\documentclass[conference]{IEEEtran}
\IEEEoverridecommandlockouts

\usepackage[utf8]{inputenc}
\usepackage[T1]{fontenc}

\usepackage{amsmath,amssymb,amsfonts}
\usepackage{algorithm}
\usepackage{algorithmic}

\usepackage{graphicx}
\usepackage{xcolor}
\usepackage{booktabs}
\usepackage{multirow}
\usepackage{array}
\usepackage{tabularx}
\usepackage{threeparttable}
\usepackage{pifont}
\usepackage{stfloats}

\usepackage{tcolorbox}
\tcbuselibrary{skins,breakable}
\usepackage{mdframed}
\usepackage{microtype}

\usepackage{cite}
\usepackage{hyperref}
\hypersetup{
  colorlinks=true,
  linkcolor=blue,
  citecolor=blue,
  urlcolor=blue
}

\providecommand{\ket}[1]{|#1\rangle}
\providecommand{\bra}[1]{\langle #1|}
\newcommand{\rhotgt}{\rho_{\mathrm{tgt}}}
\newcommand{\rhoLS}{\rho_{\mathrm{LS}}}
\newcommand{\rhoSSP}{\rho_{\mathrm{SSP}}}
\newcommand{\Fid}{\mathcal{F}}
\newcommand{\tr}{\mathrm{tr}}
\newcommand{\Var}{\mathrm{Var}}

\begin{document}

\title{SSP-QST: Spectral Subspace Purification\\
for Photonic Quantum State Tomography}

\author{
\IEEEauthorblockN{Anuvab Sen}
\IEEEauthorblockA{
\textit{School of Electrical and Computer Engineering}\\
\textit{Georgia Institute of Technology}\\
Atlanta, GA 30332, USA\\
asen74@gatech.edu}
\and
\IEEEauthorblockN{Saibal Mukhopadhyay}
\IEEEauthorblockA{
\textit{School of Electrical and Computer Engineering}\\
\textit{Georgia Institute of Technology}\\
Atlanta, GA 30332, USA\\
saibal.mukhopadhyay@ece.gatech.edu}
}

\maketitle

\begin{abstract}
Photonic quantum sensing often uses low-rank entangled probes such as Greenberger-Horne-Zeilinger (GHZ), Bell, and NOON states. Although these probes are ideally rank-1, practical quantum state tomography (QST) can produce density-matrix estimates with many small finite-shot and noise-induced eigenmodes. This eigenvalue contamination can increase the estimated entropy of the reconstruction and reduce the quantum Fisher information (QFI) available for downstream sensing, while fixed rank-1 purification can discard valid signal modes when real probes acquire additional signal modes. We introduce \emph{Spectral Subspace Purification for Quantum State Tomography} (SSP-QST), a rank-adaptive post-processing layer for least-squares quantum state tomography (LS-QST). SSP-QST eigendecomposes the least-squares estimate, computes a Weyl-motivated noise floor from the measured spectrum and shot count, removes eigenmodes below this floor, and renormalises the retained subspace. It requires no rank prior, no iterative optimisation, and only one eigendecomposition. In Qiskit Aer simulations, SSP-QST achieves the highest fidelity among the tested non-iterative baselines across the evaluated probe ranks, with a maximum fidelity gain of $+0.584$. It also improves shot efficiency by at least $8\times$ within the tested range. These results show that SSP-QST can make photonic QST more reliable under finite-shot noise while providing a lightweight reconstruction primitive for feedback-oriented quantum sensing pipelines. 
\end{abstract}


\section{Introduction}
\label{sec:intro}

The promise of quantum sensing rests on a clean theoretical result.
Entangled photonic probes can estimate a physical parameter with precision
scaling as $1/n$, the Heisenberg limit, compared to the $1/\sqrt{n}$
classical bound achievable without
entanglement~\cite{giovannetti2004quantum,slussarenko2019photonic}.
Realising this advantage in practice requires not only preparing the
correct probe state but also \emph{reconstructing} it faithfully after it
has interacted with the measured system. Quantum state tomography (QST)
fulfils this role by producing a density-matrix estimate $\hat\rho$
from photon-counting measurements across multiple Pauli bases. Every
downstream parameter estimator then operates under the quantum
Cram\'{e}r-Rao bound (QCRB),
\begin{equation}
  \Var(\hat\omega) \;\geq\; \frac{1}{\nu\, F_Q[\hat\rho,\, H]},
  \label{eq:qcrb}
\end{equation}
where $F_Q[\hat\rho, H]$ is the quantum Fisher information (QFI) of the
\emph{reconstructed} state with respect to the encoding Hamiltonian $H$,
and $\nu$ is the number of independent trials~\cite{liu2020quantum}.
The implication is immediate: if tomography distorts the reconstructed
state, it can also distort the sensing precision inferred from that state.

The central issue studied in this paper is that standard tomography can inflate the eigenvalue spectrum of the reconstructed density matrix. Many photonic sensing
probes are designed to be low-rank. GHZ, Bell, and NOON
($\ket{N,0} + \ket{0,N}$) states are rank-1 in the ideal case, with
density matrix $\ket\psi\bra\psi$, and have been demonstrated across
photonic sensor platforms~\cite{xanadu2022borealis}. In practice, however, the
state reconstructed by a standard least-squares QST pipeline often contains
many small eigenvalues. Some of these eigenvalues correspond to physical
noise, while others are produced by finite-shot measurement fluctuations.
Keeping all of them produces a noise-inflated, effectively full-rank
estimate $\rhoLS$, which can raise the entropy of the reconstruction and
generally reduce the useful QFI of the estimated probe
state~\cite{toth2012multipartite}.

At the same time, simply forcing the reconstruction to be rank-1 is not a
reliable solution. Real photonic probes are often low-rank but not exactly
pure. Beam-splitter imbalance, multiphoton emission, waveguide mode
crosstalk, photon loss, and slow source drift can introduce additional
physical signal components, elevating the effective probe rank to
$r>1$~\cite{altepeter2005photonic,slussarenko2019photonic,demkowicz2015quantum,xanadu2022borealis}.
A rank-one reconstruction method is then structurally biased, even with
perfect measurements, it can only retain the dominant eigenmode and must
discard the remaining valid signal subspace. Thus, the reconstruction
problem has two opposite failure modes. Least-squares QST keeps too many
noise modes, while rank-one purification keeps too few signal modes. The
goal of this work is to recover the low-rank signal subspace without
knowing its rank in advance.

Prior work has recognised that low-rank structure is exploitable in
quantum tomography~\cite{cramer2010efficient}. Compressed sensing
QST~\cite{gross2010quantum} provides rank-dependent sample-complexity
guarantees and can recover low-rank states from fewer measurements, but
practical recovery typically relies on iterative convex or regularised
optimisation, making it less suitable for low-latency feedback loops.
Iterative maximum-likelihood estimation
(MLE)~\cite{hradil1997quantum,rehacek2007diluted,wu2024fast}
can produce statistically well-behaved estimates, but requires repeated
matrix updates and eigendecompositions. Hedged
MLE~\cite{blumekohout2010hedged} improves predictive performance by
regularising the estimate toward the maximally mixed state, but this also
biases the spectrum toward higher rank. Linear regression estimation
(LRE)~\cite{qi2013linear,qi2017adaptive,li2026efficient} provides a fast
LS-style fit, but does not select the reconstruction rank from the data.
Top-eigenvector extraction~\cite{greenlab2022robust} is effective when the
target is known to be rank-1, but becomes structurally mismatched for
rank-$r>1$ probes. Threshold quantum state tomography
~\cite{caruccio2025threshold} adaptively prunes measurement settings, but
does not directly remove rank inflation in the reconstructed density
matrix. These methods leave open the need for a lightweight,
rank-adaptive post-processor that operates directly on an LS-QST estimate.

We propose \emph{Spectral Subspace Purification QST} (SSP-QST), a
data-driven post-processing layer for LS-QST. The method is simple:
eigendecompose $\rhoLS$, estimate a noise floor from the dominant
eigenvalue and the shot count, discard eigenvalues below that floor, and
renormalise the remaining spectrum. The threshold is motivated by the Weyl
matrix perturbation theorem, which bounds how far the eigenvalues of the
noisy reconstruction can move from those of the underlying state. For an
effectively pure probe, SSP-QST reduces to top-eigenvector purification.
For a mixed low-rank probe, it preserves all eigenmodes that rise above
the estimated noise floor. In this way, SSP-QST takes a middle path
between full-rank LS-QST and fixed rank-one purification.

The key contributions of this work are as follows.
\begin{itemize}
  \item \textbf{Data-driven rank selection for LS-QST:}
  We introduce SSP-QST, a closed-form post-processing layer that selects
  the reconstruction rank directly from the eigenspectrum of the LS-QST
  estimate. The method uses a Weyl-perturbation-motivated noise floor
  computed from the dominant eigenvalue and the shot count, requiring no
  prior knowledge of the true rank.

  \item \textbf{One reconstruction rule for pure and mixed low-rank probes:}
  SSP-QST avoids the two common failure modes of non-adaptive
  reconstruction. Unlike LS-QST, it does not retain all noise-inflated
  eigenmodes. Unlike rank-one purification, it does not discard valid
  signal modes when the photonic probe has effective rank $r>1$.

  \item \textbf{Fidelity-gain analysis and simulation validation:}
  We provide a perturbative fidelity-gain analysis showing how removing
  non-signal eigenmodes improves the reconstructed state under eigengap
  and alignment assumptions. In Qiskit Aer simulations, SSP-QST achieves
  the highest fidelity across all tested probe ranks and improves photon
  efficiency within the tested shot-budget range.

  \item \textbf{Low-latency post-processing for feedback-oriented sensing:}
  SSP-QST requires only one eigendecomposition and renormalisation, giving
  an $O(d^3)$ closed-form reconstruction. This makes it suitable as a
  lightweight reconstruction primitive inside feedback-oriented photonic
  sensing pipelines. Recent FPGA-based QST implementations support the
  feasibility of deploying comparable reconstruction workloads on
  reconfigurable hardware pipelines~\cite{miller2023fpga,wu2025fpga}.
\end{itemize}

\section{Background and Related Work}
\label{sec:bg}

This section establishes the formalism of photonic QST, the connection
between eigenrank and sensing precision, the physical sources of rank
elevation, and the limitations of prior approaches.

\textbf{Photonic quantum state tomography :}
An $n$-qubit photonic state is described by a density operator
$\rho \in \mathbb{C}^{d \times d}$ with $d = 2^n$,
$\rho \succeq 0$, $\tr(\rho) = 1$. Full
tomography~\cite{altepeter2005photonic,paris2004quantum} measures the $4^n - 1$
non-identity Pauli operators $\{P_k\}$; each corresponds physically
to a distinct wave-plate configuration before a polarising beam splitter.
Under $N_s$-shot photon-coincidence statistics, the measured expectations
satisfy
\begin{equation}
  e_k = \tr(P_k \rho) + \eta_k, \quad
  \eta_k \sim \mathcal{N}\!\!\left(0,\,
    \frac{1 - [\tr(P_k\rho)]^2}{N_s}\right)
  \label{eq:meas}
\end{equation}
and the standard least-squares reconstruction is
\begin{equation}
  \rhoLS = \Pi_{\mathrm{phys}}\!\!\left[
    \frac{I_d}{d} + \frac{1}{d}\sum_k e_k P_k
  \right]
  \label{eq:ls}
\end{equation}
where $\Pi_{\mathrm{phys}}$ projects onto the set of valid density
operators by clipping negative eigenvalues and
renormalising~\cite{smolin2012efficient}. Under any
non-zero noise, $\rhoLS$ has full eigenrank $d$ regardless of the true
state rank, and this is the failure mode SSP-QST corrects.

\textbf{Quantum Fisher information and sensing precision :}
For state $\rho = \sum_i \lambda_i \ket{i}\bra{i}$ and the
phase-sensing generator $H = J_z = \frac{1}{2}\sum_q Z_q$:
\begin{equation}
  F_Q[\rho, H] = 2 \sum_{\substack{i,j \\ \lambda_i + \lambda_j > 0}}
  \frac{(\lambda_i - \lambda_j)^2}{\lambda_i + \lambda_j}
  \left|\bra{i} H \ket{j}\right|^2
  \label{eq:qfi}
\end{equation}
The Heisenberg limit for an $n$-qubit GHZ probe is
$F_Q^{\mathrm{max}} = n^2$. Because QFI is convex in
$\rho$~\cite{toth2012multipartite}, any noise-induced mixing of the state
strictly reduces $F_Q$. Rank inflation in $\rhoLS$ distributes eigenvalue
weight across spurious noise modes and diminishes the contrasts
$(\lambda_i - \lambda_j)^2 / (\lambda_i + \lambda_j)$ that drive $F_Q$.
SSP-QST recovers the low-rank structure and partially restores those
contrasts in the reconstructed estimate.

\textbf{Hardware errors and probe rank elevation :}
Ideal entangled probe states are rank-1 pure states, but hardware errors
systematically elevate the effective eigenrank of the prepared state.
Beam-splitter reflectivity errors cause a nominally rank-1 GHZ state to
emerge as a rank-2 mixture; spontaneous parametric down-conversion (SPDC) sources
emit higher-order Fock components through simultaneous multi-pair events,
increasing the eigenrank beyond one; evanescent coupling errors between
adjacent waveguides in integrated silicon-photonic chips mix spatial modes,
producing states whose eigenrank grows with the number of coupled guides;
and source parameter drift in adaptive sensing protocols continuously
displaces the prepared state from its nominal rank-1 target. In every case,
the effective eigenrank $r$ is unknown at reconstruction time.
Rank-constrained methods that assume $r = 1$ introduce a systematic bias
that persists regardless of measurement quality or shot count. SSP-QST
infers $r$ directly from the eigenspectrum of $\rhoLS$ via a closed-form
threshold derived from the Weyl matrix perturbation theorem, with no
prior, no iterative procedure, and no external reference state.

\begin{figure*}[!t]
    \centering
    \includegraphics[width=0.94\textwidth]{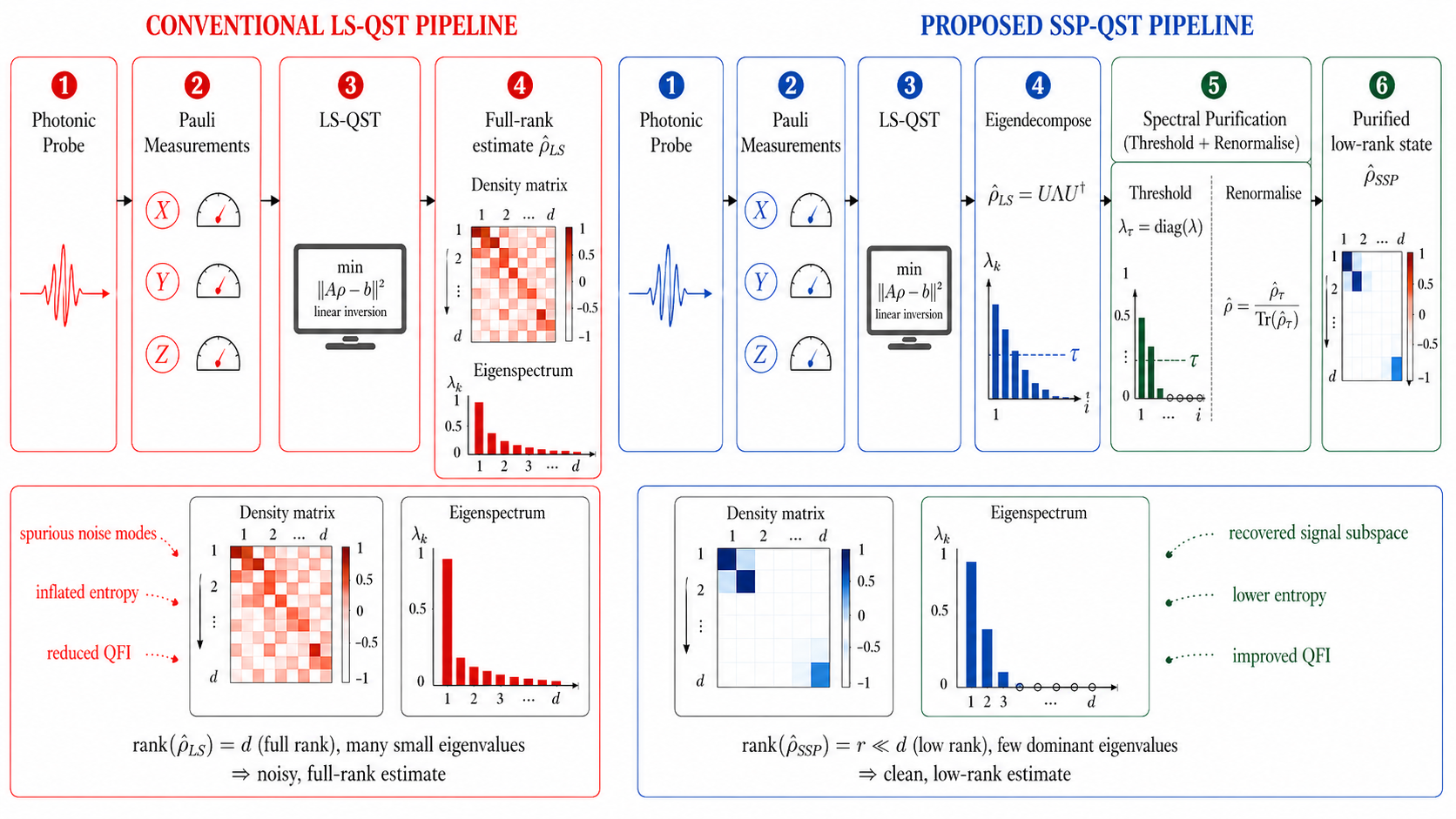}
    \caption{
    \textbf{Existing Pipelines vs.\ SSP-QST.}
    \textbf{Left:} Conventional tomography reconstructs a photonic probe from Pauli measurements using LS-QST, producing a full-rank estimate $\hat{\rho}$ with noise-inflated eigenmodes, increased entropy, and reduced quantum Fisher information (QFI).
    \textbf{Right:} SSP-QST applies eigendecomposition, spectral thresholding, and renormalisation to the same LS-QST output, removing low-eigenvalue noise modes and recovering a low-rank purified estimate $\rho_{\mathrm{SSP}}$ with improved QFI.
    }
    \label{fig:ssp_pipeline}
\end{figure*}   

\textbf{Limitations of prior rank-aware methods :}
Least-squares QST~\cite{qi2013linear} is computationally efficient
($O(d^2)$) and feedback-compatible, but applies no eigenrank constraint
and accepts all $d$ eigenmodes of $\rhoLS$ as physical signal. Compressed
sensing QST~\cite{gross2010quantum,flammia2012quantum} recovers
rank-$r$ states from $O(rd\log^2 d)$ random Pauli measurements via
nuclear norm minimisation, but requires $r$ as a prior and performs
iterative convex optimisation incompatible with real-time feedback.
Projected least-squares (PLS) tomography~\cite{guta2020fast,
surawy2022projected} achieves provable optimal sample complexity by
projecting $\rhoLS$ onto the positive semi-definite cone, but its
projection is rank-agnostic and retains all eigenmodes weighted by
truncation, providing no automatic rank identification. Iterative maximum-likelihood
estimation, including the original Hradil
formulation~\cite{hradil1997quantum} and its diluted
variant~\cite{rehacek2007diluted}, converges to the ML estimate at the
cost of multiple eigendecompositions per reconstruction. Streaming-data quantum reconstruction approaches~\cite{streaming2025}
handle non-stationary sources but introduce algorithmic latency. Hedged
MLE~\cite{blumekohout2010hedged} adds a Lidstone-style regulariser that
keeps the estimate strictly positive, but biases the spectrum toward
higher rank. Top-eigenvector
extraction~\cite{greenlab2022robust} retains only the dominant eigenvector
of $\rhoLS$ and is therefore rank-1 by construction, structurally unable
to represent rank-$r > 1$ states regardless of noise. Spectral squaring
($\rhoLS^2 / \tr(\rhoLS^2)$) sharpens the eigenspectrum by amplifying the
dominant eigenvalue and is highly effective for rank-1 targets, but for
rank-$r > 1$ states the squaring operation distorts the relative weights
of the $r$ signal eigenmodes by amplifying $\lambda_{\mathrm{max}}^2$
disproportionately. Threshold QST~\cite{caruccio2025threshold} prunes
\emph{measurement settings} adaptively from the diagonal of $\rho$ but
does not address rank inflation in the reconstructed density matrix itself.
Bayesian QST~\cite{blume2010optimal} encodes rank priors through Monte
Carlo posterior sampling at $O(M d^4)$ cost, prohibitive for real-time
feedback. Neural-network state tomography~\cite{torlai2018neural} learns
a generative model of the wavefunction from measurement data and scales
favourably for low-entanglement states, but requires offline training
and is not yet positioned for closed-loop integration. We further note that pure-state-reporting estimators including
MLE and constrained least squares are inadmissible under fidelity loss
for general states~\cite{ferrie2018inadmissible,chen2025robustness}, motivating the search
for spectrum-aware alternatives.

\textbf{Perturbative and spectral-thresholding approaches to quantum estimation :}
Matrix perturbation theory in state estimation predates this work.
Ban\'a\v{s}, \v{R}eh\'a\v{c}ek, and
Hradil~\cite{banas2006perturbative} correct eigenvalues and eigenvectors
perturbatively to accelerate iterative maximum-likelihood estimation
(broader ML frameworks are surveyed
in~\cite{banaszek1999maximum,james2001measurement,lvovsky2009continuous});
SSP-QST differs in objective and mechanism, using the Weyl theorem
non-iteratively as a closed-form criterion separating signal from noise
eigenvalues in a least-squares reconstruction. Thresholded LS
tomography~\cite{butucea2015spectral,acharya2019comparative} truncates
below a universal concentration-derived threshold with asymptotic
minimax guarantees; SSP-QST instead calibrates its threshold from the
measured spectrum ($\hat{p}$ of Eq.~\eqref{eq:thresh} plus an explicit
finite-shot margin), targeting the low-latency, feedback-oriented
photonic setting. We view SSP-QST as a new application of this
perturbative lineage, spectral purification of least-squares photonic
QST, rather than the introduction of perturbation theory to tomography.

Table~\ref{tab:related} summarises this landscape. SSP-QST is the only
method that combines closed-form rank adaptivity without a prior, an
$O(d^3)$ procedure, and compatibility with real-time closed-loop feedback.

\begin{table}[t]
\centering
\caption{Qualitative Comparison of QST Methods}
\label{tab:related}
\begin{threeparttable}
\footnotesize
\renewcommand{\arraystretch}{1.06}
\setlength{\tabcolsep}{3.4pt}

\begin{tabular}{@{}lcccc@{}}
\toprule
\textbf{Method}
& \textbf{Rank-Adapt.}
& \textbf{Prior?}
& \textbf{Cost}
& \textbf{RT FB} \\
\midrule

LS-QST~\cite{qi2013linear}
& No & None & $O(d^2)$ & \checkmark \\

Spec.\ Squaring
& No & Rank 1 & $O(d^2)$ & Limited \\

Top Eigvec.~\cite{greenlab2022robust}
& No & Rank 1 & $O(d^3)$ & Limited \\

Iterative MLE~\cite{rehacek2007diluted}
& No & None & $O(Kd^3)$ & No \\

Hedged MLE~\cite{blumekohout2010hedged}
& No & None & $O(Kd^3)$ & No \\

Comp.\ Sens.~\cite{gross2010quantum}
& Yes & Rank $r$ & Iter.\ SDP\tnote{$\dagger$} & No \\

Bayesian~\cite{blume2010optimal}
& Yes & Model & $O(Md^4)$ & No \\

\midrule

\textbf{SSP-QST (Ours)}
& \textbf{Yes}
& \textbf{None}
& $\mathbf{O(d^3)}$
& \textbf{\checkmark} \\

\bottomrule
\end{tabular}

\vspace{2pt}
\begin{tablenotes}[flushleft]
\scriptsize
\item \textbf{Cost} denotes per-reconstruction computational cost, not sample complexity.
\item[$\dagger$] Compressed sensing requires $O(rd\log^2 d)$ Pauli measurements for rank-$r$ recovery, but reconstruction uses an iterative SDP and is generally unsuitable for real-time feedback.
\item \textbf{RT FB} denotes real-time feedback before photonic-source drift accumulates~\cite{altepeter2005photonic}. $K$ is the number of MLE iterations; $M$ is the number of Monte Carlo samples.
\end{tablenotes}

\end{threeparttable}
\end{table}

\section{Proposed Method}
\label{sec:method}

Fig.~\ref{fig:ssp_pipeline} contrasts the conventional LS-QST pipeline with the SSP-QST extension. SSP-QST comprises two composable components. The first is a spectral
purification step that operates on any LS-QST output and produces a
physically valid, rank-adaptive density matrix by discarding eigenmodes
below a closed-form noise floor. The second is a parameter-shift feedback
controller that uses the purified estimate as a fidelity gradient signal
to correct coherent source drift. Both components are modular, and the
purification layer deploys as a drop-in post-processor for any existing
pipeline. This section derives the threshold, presents the algorithm,
states the perturbative fidelity bound, and describes the controller.

\subsection{Noise-Floor Threshold via the Weyl Perturbation Theorem}

We derive the threshold from a classical result in matrix
analysis~\cite{bhatia1997matrix}.
We index eigenvalues in ascending order throughout, so $\lambda_d$ denotes the largest and $\lambda_{d-1}$ the second-largest.
Let $\rhoLS = \rhotgt + \Delta$, where $\Delta$ encodes the combined
effect of photonic noise and finite-shot measurement error. The Weyl
theorem guarantees:
\begin{equation}
  |\lambda_i(\rhoLS) - \lambda_i(\rhotgt)| \;\leq\; \|\Delta\|_2
  \label{eq:weyl}
\end{equation}
for all $i$. For a rank-1 target under depolarising noise $p$, the
dominant eigenvalue satisfies $\lambda_d(\rhoLS) \approx 1 - p(d-1)/d$,
so the deviation $\hat{p} = 1 - \lambda_d \approx p(d-1)/d$ directly
estimates the noise level from the reconstruction itself. Noise
eigenmodes scale as $p/d = \hat{p}/(d-1)$; adding a half-standard-deviation
shot-noise margin yields the threshold:
\begin{equation}
  \epsilon_{\mathrm{th}} = \frac{\hat{p}}{d - 1} +
  \frac{0.5}{\sqrt{N_s}},
  \qquad \hat{p} \;=\; \max\!\bigl(0,\; 1 - \lambda_d\bigr)
  \label{eq:thresh}
\end{equation}
Eigenvalues above $\epsilon_{\mathrm{th}}$ are classified as signal;
those below are noise and are discarded. Correct rank identification
requires an eigengap condition: the smallest true signal eigenvalue
$\lambda_r^{\mathrm{tgt}}$ must satisfy
$\lambda_r^{\mathrm{tgt}} > \epsilon_{\mathrm{th}} + \|\Delta\|_2$,
so that Weyl's bound (Eq.~\ref{eq:weyl}) places the perturbed signal
eigenvalue strictly above the noise floor. Under this separation,
all $r$ signal eigenvalues lie above $\epsilon_{\mathrm{th}}$ and
SSP-QST recovers the rank automatically. When the separation is
violated, the algorithm degrades gracefully by underestimating the rank,
the safe failure mode in which signal modes are never falsely added
(this behaviour is documented in Section~\ref{sec:results}).

We emphasise the scope of this derivation. The noise-eigenvalue scaling
$p/d$ that fixes the first term of Eq.~\eqref{eq:thresh} is exact for the
depolarising channel, whose isotropic perturbation
$(p/d)(I_d - \rhotgt)$ distributes weight uniformly across the noise
subspace. Under amplitude damping or phase damping, the perturbation
$\Delta$ is structured rather than isotropic, and its eigenvalues need not
concentrate near $p/d$; for those channels, Eq.~\eqref{eq:thresh} is a
heuristic motivated by the Weyl scale rather than a rigorously derived
bound. Its use beyond depolarising noise is therefore justified
empirically: Sections~\ref{subsec:noise_robustness}
and~\ref{subsec:additional_validation} show that fidelity and rank
identification are preserved under amplitude- and phase-damping channels
in simulation. For local amplitude damping the envelope is provable:
telescoping, $\lVert \cdot \rVert_{\mathrm{op}} \le \lVert \cdot
\rVert_1$, and the elementary bound $\lVert \Lambda_\gamma -
\mathrm{id} \rVert_\diamond \le \gamma + 2a + a^2 \le 2\gamma(1+\gamma)$,
$a = 1{-}\sqrt{1{-}\gamma}$ (numerically ${=}\,2\gamma$), give
$\lVert \Delta \rVert_{\mathrm{op}} \le 2n\gamma(1+\gamma)$ via Weyl;
observed shifts stay within $n\gamma$ (largest $0.64\,n\gamma$ at
$n{=}4$, $\gamma{=}0.06$). This bounds the envelope, not the floor
location; a tight structured-noise derivation is future work.

The shot-noise margin $0.5/\sqrt{N_s}$ admits a concentration
justification. With $\Delta_{\mathrm{shot}} = \frac{1}{d}\sum_{P \neq I}
\delta_P P$ and $\mathrm{Var}(\delta_P) \le 1/N_s$, first-order
perturbation theory gives
$\delta\lambda_i \approx \frac{1}{d}\sum_P \delta_P \langle v_i|P|v_i
\rangle$; Pauli completeness ($\sum_P \langle v|P|v \rangle^2 = d$)
yields $\mathrm{Var}(\delta\lambda_i) \le 1/(dN_s)$, so the margin is
$\sqrt{d}/2$ standard deviations. Hoeffding sub-Gaussianity then gives
$\Pr[\delta\lambda_i > 0.5/\sqrt{N_s}] \le e^{-d/8}$ ($\le e^{-2}$
already at $n{=}4$), decaying exponentially in dimension; with the
separation from the first term of Eq.~\eqref{eq:thresh} and graceful
degradation, this accounts for the rank-identification accuracy of
Section~\ref{subsec:rank_threshold}.

A dimension-aware complement: the summands $\delta_P P/d$ are
independent, zero-mean, and bounded, so matrix
Bernstein~\cite{tropp2015introduction} gives $\lVert
\Delta_{\mathrm{shot}} \rVert_2 \lesssim \sqrt{2\ln(2d)/N_s}$ whp
($0.041$ here), bounding every eigenvalue shift via Weyl. This worst case is deliberately not the margin: substituting
it in Eq.~\eqref{eq:thresh} inflates the threshold ${\sim}5\times$ and
over-truncates (rank-7 fidelity $0.941 \to 0.786$). The margin tracks
the typical deviation; Bernstein is the rigorous uniform fallback.

\subsection{Spectral Subspace Purification Algorithm}

\begin{figure*}[t]
\begin{tcolorbox}[
  enhanced, colback=white, colframe=black!72, boxrule=0.7pt, arc=3pt,
  title={\textbf{Algorithm 1:\enspace Spectral Subspace Purification QST (SSP-QST)}},
  fonttitle=\bfseries\small, coltitle=white,
  attach boxed title to top left={yshift=-2.2mm, xshift=5mm},
  boxed title style={colback=black!82, arc=2pt, boxrule=0pt, left=3pt, right=3pt},
  top=4mm, bottom=3mm, left=4mm, right=4mm, width=\textwidth,
]
\small\setlength{\parskip}{0pt}
\noindent
\begin{minipage}[t]{0.45\textwidth}
  \textbf{Input:}\enspace
  LS estimate $\rhoLS \in \mathbb{C}^{d \times d}$;\enspace shots $N_s$
\end{minipage}%
\hfill
\begin{minipage}[t]{0.52\textwidth}
  \textbf{Output:}\enspace
  Purified state $\rhoSSP$
  \;$(\rhoSSP \succeq 0,\;\tr(\rhoSSP) = 1)$
\end{minipage}

\smallskip\hrule\smallskip

\begin{minipage}[t]{0.48\textwidth}
\begin{algorithmic}[1]
\setcounter{ALC@line}{0}
\STATE $\rhoLS \leftarrow \tfrac{1}{2}(\rhoLS + \rhoLS^\dagger)$
  \hfill\textit{\small $\triangleright$ Symmetrise}
\STATE $\{\lambda_i, \ket{v_i}\}_{i=1}^{d}
  \leftarrow \mathrm{eigh}(\rhoLS)$
  \hfill\textit{\small $\triangleright$ Hermitian eigendecomposition}
\STATE $\lambda_i \leftarrow \max(\lambda_i, 0)\;\forall i$
  \hfill\textit{\small $\triangleright$ Enforce positivity}
\STATE $\hat{p} \leftarrow \max(0,\; 1 - \lambda_d)$
  \hfill\textit{\small $\triangleright$ Estimate noise level}
\STATE $\epsilon_{\mathrm{th}} \leftarrow
  \dfrac{\hat{p}}{d-1} + \dfrac{0.5}{\sqrt{N_s}}$
  \hfill\textit{\small $\triangleright$ Weyl noise floor}
\end{algorithmic}
\end{minipage}%
\hfill\vrule\hfill
\begin{minipage}[t]{0.46\textwidth}
\begin{algorithmic}[1]
\setcounter{ALC@line}{5}
\IF{$\lambda_{d-1} < 2\,\epsilon_{\mathrm{th}}$}
  \RETURN $\rhoSSP = \ket{v_d}\!\bra{v_d}$
  \hfill\textit{\small $\triangleright$ Rank-1 override}
\ENDIF
\STATE $\tilde{\lambda}_i \leftarrow
  \lambda_i \cdot \mathbf{1}[\lambda_i > \epsilon_{\mathrm{th}}]\;\forall i$
  \hfill\textit{\small $\triangleright$ Threshold signal modes}
\STATE $\hat{\lambda}_i \leftarrow
  \tilde{\lambda}_i \big/ \sum_j \tilde{\lambda}_j\;\forall i$
  \hfill\textit{\small $\triangleright$ Renormalise: $\tr(\rhoSSP)=1$}
\RETURN $\rhoSSP =
  \displaystyle\sum_{i=1}^{d} \hat{\lambda}_i\, \ket{v_i}\!\bra{v_i}$
\end{algorithmic}
\end{minipage}
\end{tcolorbox}
\label{alg:ssp}
\end{figure*}

Algorithm~1 produces a valid density matrix at every step:
$\rhoSSP \succeq 0$ (all $\hat\lambda_i \geq 0$),
$\tr(\rhoSSP) = 1$ (renormalisation), and $\rhoSSP = \rhoSSP^\dagger$.
The dominant computational cost is the eigendecomposition at Step 2,
$O(d^3)$, and completes in sub-microsecond time on any modern processor
for $n = 4$ ($d = 16$).

\subsection{Perturbative Fidelity Bound}

\begin{mdframed}[backgroundcolor=blue!3,linecolor=blue!20,
  innertopmargin=6pt,innerbottommargin=6pt,
  innerleftmargin=8pt,innerrightmargin=8pt,
  userdefinedwidth=\columnwidth]
\noindent\textbf{Proposition 1.}\;
Let $\rhotgt = \ket\psi\bra\psi$ be a pure target state, and let
$\rhoLS$ admit the spectral decomposition
$\rhoLS = \sum_{i=1}^{d} \lambda_i \ket{v_i}\!\bra{v_i}$ with eigenvalues
ordered as $\lambda_d \geq \lambda_{d-1} \geq \cdots \geq \lambda_1$.
Suppose the threshold $\epsilon_{\mathrm{th}}$ satisfies
$\hat{p}/(d-1) < \epsilon_{\mathrm{th}} < \lambda_d$, so that only the
dominant eigenvector survives, and assume the alignment condition
$|\langle v_d \,|\, \psi \rangle| \approx 1$. Then
\begin{equation}
  \Fid(\rhoSSP, \rhotgt) - \Fid(\rhoLS, \rhotgt)
  \;\approx\;
  \frac{p(d-1)}{d}\,|\langle v_d\,|\,\psi\rangle|^{2}
  \;\geq\; 0.
  \label{eq:gain}
\end{equation}
\end{mdframed}

\smallskip
\noindent\textbf{\textit{Proof.}}\;
By the Uhlmann fidelity~\cite{uhlmann1976transition},
\begin{equation}
  \Fid(\rhoSSP, \rhotgt) \;=\; |\langle v_d \,|\, \psi \rangle|^{2},
  \label{eq:F_ssp}
\end{equation}
\begin{equation}
  \Fid(\rhoLS, \rhotgt)
  \;=\; \lambda_d\,|\langle v_d \,|\, \psi \rangle|^{2}
   + \sum_{i<d} \lambda_i\,|\langle v_i \,|\, \psi \rangle|^{2}.
  \label{eq:F_ls}
\end{equation}
Subtracting Eq.~\eqref{eq:F_ls} from Eq.~\eqref{eq:F_ssp} and
substituting the leading-order expansion
$\lambda_d \approx 1 - p(d-1)/d$ yields Eq.~\eqref{eq:gain}.

\smallskip
The bound scales as $p(d-1)/d$, growing with both the noise strength
$p$ and the Hilbert-space dimension $d = 2^{n}$. The signal-to-noise
eigenvalue separation therefore widens with system size, so SSP-QST's
advantage strengthens as photonic systems scale.

\smallskip
\noindent\textbf{\textit{Multi-rank extension :}}\;
The same argument generalises to rank-$r > 1$ targets. Let
$\rhotgt = \sum_{j=1}^{r} w_j \ket{\phi_j}\!\bra{\phi_j}$ with
$\sum_{j=1}^{r} w_j = 1$ and orthonormal $\{\ket{\phi_j}\}$, and assume
the eigengap condition $\lambda_r^{\mathrm{tgt}} > \epsilon_{\mathrm{th}}
+ \|\Delta\|_{2}$, so that all $r$ signal eigenvalues of $\rhoLS$ survive
thresholding. Under the alignment condition
$|\langle v_{d-j+1} \,|\, \phi_j \rangle| \approx 1$ for
$j = 1, \ldots, r$, the SSP-QST estimate becomes
\begin{equation}
  \rhoSSP \;=\; \sum_{j=1}^{r}
  \tilde{w}_j \ket{v_{d-j+1}}\!\bra{v_{d-j+1}},
  \;\;
  \tilde{w}_j \;=\; \frac{\lambda_{d-j+1}}
  {\sum_{k=1}^{r}\lambda_{d-k+1}},
  \label{eq:rho_ssp_multi}
\end{equation}
where renormalisation removes the $d-r$ noise eigenmodes that $\rhoLS$
retains. Repeating the Uhlmann fidelity calculation gives
\begin{equation}
\begin{aligned}
\Delta F
&\triangleq
\Fid(\rhoSSP,\rhotgt)-\Fid(\rhoLS,\rhotgt) \\[-0.5mm]
&\approx
\frac{p(d-r)}{d}
\sum_{j=1}^{r} w_j
\bigl|\langle v_{d-j+1}\,|\,\phi_j\rangle\bigr|^{2}
\geq 0 .
\end{aligned}
\label{eq:gain_multi}
\end{equation}
which recovers Eq.~\eqref{eq:gain} when $r = 1$. The gain scales as
$p(d-r)/d$: it remains strictly positive whenever the probe rank is
below the Hilbert-space dimension and weakens monotonically as $r$
approaches $d$, consistent with the rank-7 row of
Table~\ref{tab:vs_rank}, where the relative gain begins to compress.
SSP-QST's advantage is therefore largest in the low-rank regime
($r \ll d$) that characterises every realistic photonic probe.

\begin{figure*}[t]
  \centering
  \includegraphics[
    width=0.95\textwidth,
    trim={0cm 0cm 0cm 2.5cm},
    clip
  ]{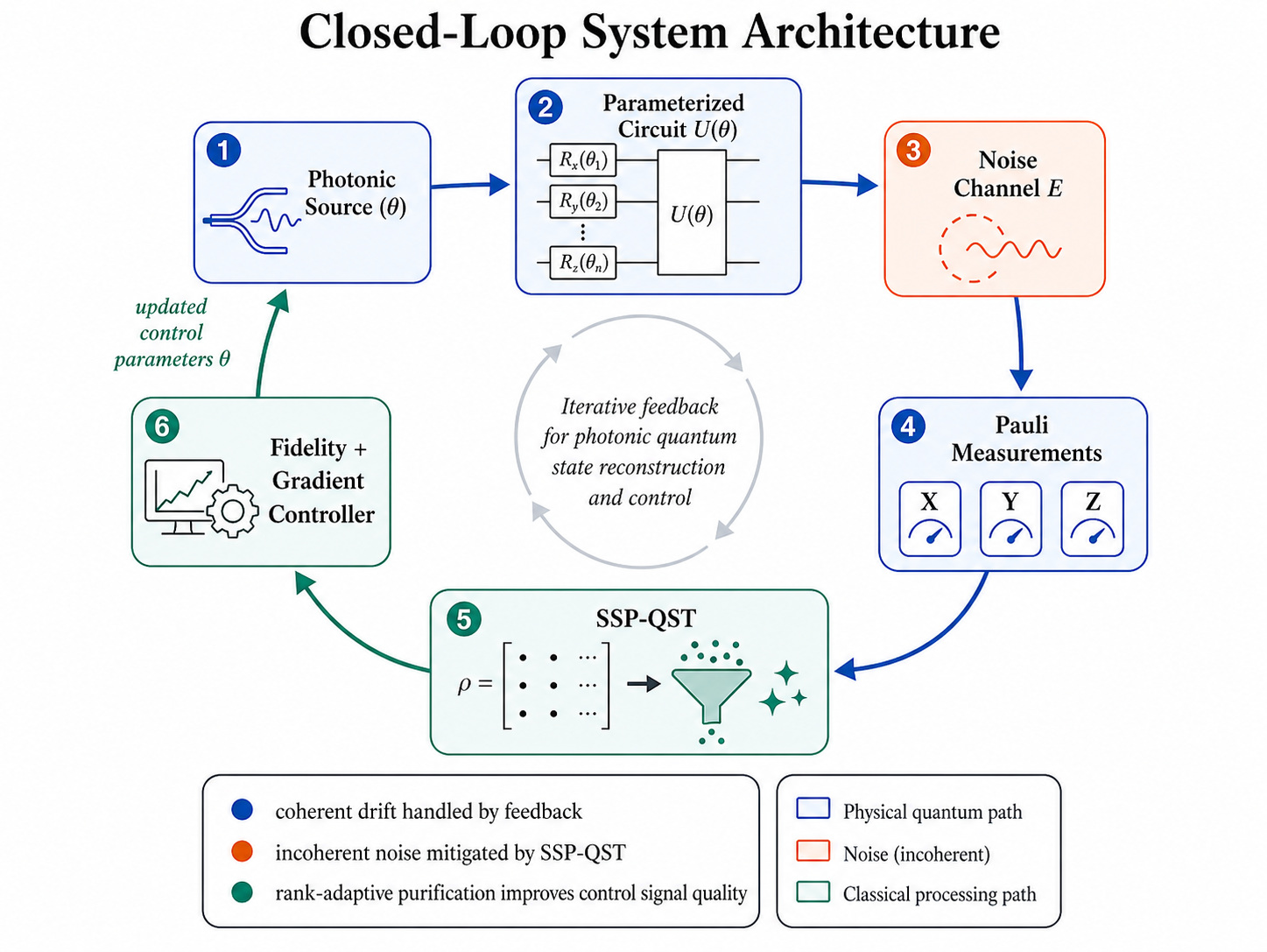}
\caption{\textbf{Closed-loop system architecture.}
A parameterised photonic source prepares a probe state, which is corrupted
by the noise channel $\mathcal{E}$ and measured across Pauli bases.
The classical reconstruction module applies LS-QST followed by SSP-QST,
then computes a parameter-shift fidelity gradient to update
$\boldsymbol{\theta}$.}
  \label{fig:closed_loop_architecture}
\end{figure*}

\subsection{Physical Parameter Feedback Controller}

Fig.~\ref{fig:closed_loop_architecture} shows the full closed-loop architecture. Photonic source imperfections fall into two categories that require
different corrective mechanisms. Coherent drift, arising from wave-plate
creep, electro-optic modulator (EOM) voltage offsets, and slow thermal phase shifts, displaces the
control vector $\boldsymbol\theta$ from its calibrated operating point and
can be corrected by parameter adjustment. Incoherent noise, arising from
photon loss and dephasing, randomises the probe state stochastically in
each shot and cannot be undone by any classical post-processing. SSP-QST
addresses both: it mitigates the reconstruction error induced by incoherent noise, and provides the fidelity gradient signal that the feedback
controller uses to correct coherent drift.

We model the photonic source as a parameterised circuit
$\mathcal{U}(\boldsymbol\theta)$, namely a GHZ preparation followed by a
per-qubit $R_y(\theta_i)$ rotation layer representing adjustable
wave-plate corrections. We recover the calibrated point by ascending
the SSP-QST fidelity:
\begin{equation}
  \boldsymbol\theta^{(t+1)} = \boldsymbol\theta^{(t)}
  + \eta \,\nabla_{\!\boldsymbol\theta}\,
  \Fid\!\bigl(\rhoSSP(\boldsymbol\theta^{(t)}),\, \rhotgt\bigr)
  \label{eq:update}
\end{equation}
The gradient is estimated by the parameter-shift rule, exact for
$R_y$-family gates:
\begin{equation}
  \frac{\partial \Fid}{\partial \theta_i}
  = \frac{1}{2}\Bigl[
    \Fid\!\bigl(\boldsymbol\theta + \tfrac{\pi}{2}\mathbf{e}_i\bigr)
  - \Fid\!\bigl(\boldsymbol\theta - \tfrac{\pi}{2}\mathbf{e}_i\bigr)
  \Bigr]
  \label{eq:pshift}
\end{equation}
Using $\rhoSSP$ rather than $\rhoLS$ in
Eqs.~\eqref{eq:update} and~\eqref{eq:pshift} reduces gradient variance
by zeroing shot-noise-inflated eigenmodes before they corrupt the gradient.

\section{Experimental Evaluation}
\label{sec:results}

This section is organised into six parts: the simulation setup, an
overview of the evaluation protocol, reconstruction fidelity across probe
rank, robustness across noise levels, shot efficiency, and
rank-identification/threshold-sensitivity analyses.

\subsection{Simulation Setup}
\label{sec:setup}

All experiments use Qiskit~2.4.1 with Qiskit-Aer~0.17.2~\cite{qiskit2026}
in software simulation; no physical hardware is used or implied. The Aer
\texttt{density\_matrix} backend evolves the full $d \times d$ density
matrix exactly under the Kraus operators of the noise model; shot noise
is overlaid analytically via Eq.~\eqref{eq:meas}, replicating the
statistics of $N_s$-shot photon-coincidence measurements precisely.

Rank-$r$ probe states are constructed as orthonormal mixtures
$\rhotgt = \sum_{i=1}^{r} w_i \ket{\phi_i}\bra{\phi_i}$, with
$\{\ket{\phi_i}\}$ obtained from QR decomposition of a Gaussian random
matrix and weights $\{w_i\}$ drawn from a Dirichlet distribution.
Concretely, QR orthonormalisation of a $d \times r$ complex Ginibre
matrix yields a Haar-distributed orthonormal $r$-frame, sampling the
signal subspace unitarily invariantly over all rank-$r$ subspaces of
$\mathbb{C}^d$; weights are drawn from the flat
$\mathrm{Dir}(1,\ldots,1)$, uniform on the simplex, so no spectral
profile is privileged. This
models any photonic source whose dominant modes span a rank-$r$
subspace, physically arising from mode crosstalk, multiphoton emission,
or incomplete variational state preparation. Each rank is averaged over
8 to 10 independently drawn random targets.

Three photonic-inspired channels are applied as gate-level Kraus
errors: depolarising at rate $p$, with
$\mathcal{E}_{\mathrm{dep}}(\rho) = (1-p)\rho + (p/d)I_d$;
amplitude damping at rate $\gamma$; and phase damping at rate
$\lambda$. We compare SSP-QST against three non-iterative baselines:
LS-QST (Eq.~\ref{eq:ls}); spectral squaring
$\rhoLS^2/\tr(\rhoLS^2)$, a rank-1 purity-boosting heuristic; and
top-eigenvector extraction~\cite{greenlab2022robust}. We deliberately
restrict the head-to-head comparison to non-iterative methods because
real-time feedback is the design constraint; iterative MLE
variants~\cite{hradil1997quantum,rehacek2007diluted,blumekohout2010hedged},
compressed sensing~\cite{gross2010quantum}, and
Bayesian~\cite{blume2010optimal} reconstructions all exceed the latency
budget set by photonic source drift~\cite{altepeter2005photonic}.
All experiments use $N_s = 4096$ shots per Pauli basis element and
random seed 42.

\subsection{Overview of Evaluation}
\label{subsec:eval_overview}

We evaluate SSP-QST on rank-$r$ photonic probe states for $r=2$ through
$7$, where different non-adaptive reconstruction strategies exhibit
complementary failure modes. LS-QST retains all eigenmodes and therefore
preserves noise-inflated full-rank structure, while rank-one
post-processing retains only the dominant eigenmode and discards valid
signal components when $r>1$. Spectral squaring sharpens the spectrum but
can distort the relative weights of multiple signal eigenmodes. The evaluation focuses on three axes: rank-dependent reconstruction fidelity, robustness to increasing noise, and photon efficiency under limited shot budgets.

\subsection{Reconstruction Fidelity Across Probe Rank}
\label{subsec:rank_fidelity}

\begin{figure*}[t]
  \centering
  \includegraphics[width=0.8\textwidth]{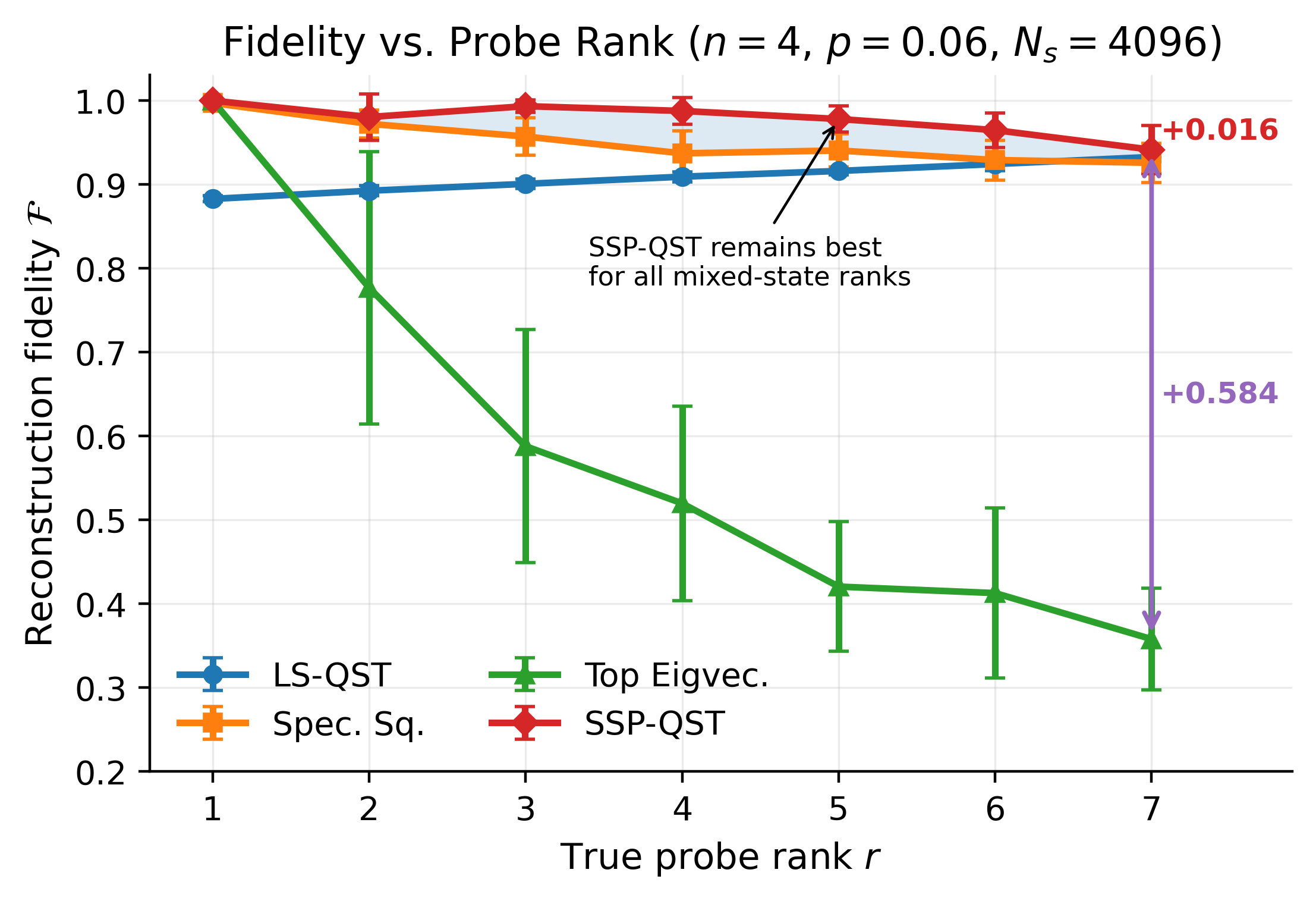}
  \caption{Reconstruction fidelity versus true probe rank $r$ for $n=4$
  qubits ($d=16$), depolarising noise $p=0.06$, $N_s=4096$ shots per
  Pauli basis. Each point is averaged over 14 independently drawn
  rank-$r$ states; error bars denote one standard deviation. The shaded
  region marks SSP-QST's advantage over spectral squaring, the strongest
  non-iterative baseline in the mixed-state regime. SSP-QST matches the
  rank-1 top-eigenvector solution for pure probes and is the best method
  for every mixed-state rank tested.}
  \label{fig:vs_rank}
\end{figure*}

\begin{table}[t]
\centering
\caption{
  Reconstruction fidelity as a function of probe rank for $n=4$ qubits
  under depolarising noise $p=0.06$ with $N_s=4096$ shots per basis.
}
\label{tab:vs_rank}
\renewcommand{\arraystretch}{1.15}
\setlength{\tabcolsep}{5.5pt}
\begin{tabular}{@{}c c c c c@{}}
\toprule
\multirow{2}{*}{Probe rank $r$}
  & \multicolumn{4}{c}{Reconstruction fidelity $\mathcal{F}$} \\
\cmidrule(l){2-5}
  & LS-QST
  & Spec.\ Sq.
  & Top eigvec.
  & \textbf{SSP-QST} \\
\midrule
1 & 0.883 & 0.997 & 1.000 & \textbf{1.000} \\
\midrule
2 & 0.892 & 0.972 & 0.777 & \textbf{0.980} \\
3 & 0.901 & 0.957 & 0.588 & \textbf{0.993} \\
4 & 0.909 & 0.937 & 0.520 & \textbf{0.987} \\
5 & 0.916 & 0.940 & 0.420 & \textbf{0.978} \\
6 & 0.924 & 0.929 & 0.412 & \textbf{0.965} \\
7 & 0.933 & 0.926 & 0.358 & \textbf{0.941} \\
\bottomrule
\end{tabular}
\end{table}

\begin{figure*}[t]
  \centering
  \includegraphics[width=\textwidth]{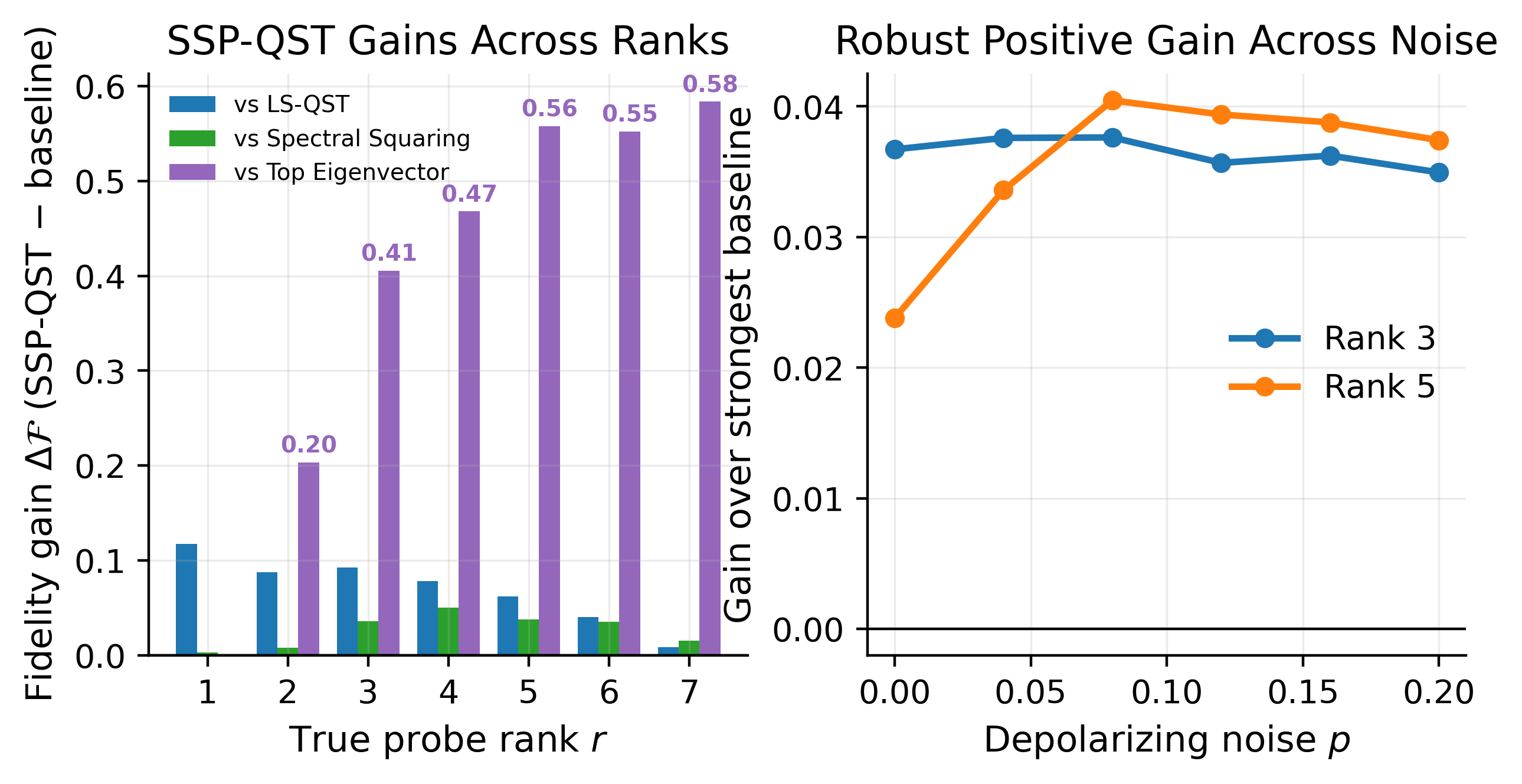}
  \caption{(a) Absolute fidelity gain $\Delta\mathcal{F}$ of SSP-QST
  over each baseline as a function of probe rank ($n=4$, $p=0.06$,
  $N_s=4096$). The gain over top-eigenvector extraction grows from
  $+0.204$ at rank-2 to $+0.584$ at rank-7. The gain over spectral
  squaring is positive across all mixed-state ranks and peaks at rank-4
  ($+0.051$). (b) Fidelity gain over the strongest baseline as a
  function of depolarising rate $p$ for rank-3 and rank-5 probes. The
  gain remains positive across the tested noise range.}
  \label{fig:gain}
\end{figure*}

Table~\ref{tab:vs_rank} and Fig.~\ref{fig:vs_rank} establish the
central quantitative result. SSP-QST achieves the highest fidelity of
any tested method at every probe rank from $r = 1$ through $r = 7$.
At rank-1, the rank-1 override (Algorithm~1, line~6) detects that the
second eigenvalue is statistically consistent with shot noise and
returns $\ket{v_d}\!\bra{v_d}$ exactly, matching top-eigenvector
extraction at $\Fid = 1.000$. From rank-2 onward, the multi-rank
threshold takes over and SSP-QST extends its advantage. The gap is
small but consistent against spectral squaring (between $+0.008$ and
$+0.051$) and substantial against top-eigenvector extraction (between
$+0.204$ and $+0.584$). LS-QST remains below SSP-QST throughout because
its full-rank output retains the noise-inflated eigenmodes that
suppress the useful low-rank structure. The mild rise in LS-QST
fidelity with rank reflects that depolarised mixtures sit closer to
higher-rank targets under fidelity, not improved estimation.
Fig.~\ref{fig:gain}(a) summarises this behaviour. SSP-QST's advantage
persists across the mixed-state regime, reflecting the increasing
mismatch between rank-1 reconstruction and the true rank-$r$ signal
subspace.

\subsection{Robustness Across Noise Levels}
\label{subsec:noise_robustness}

Fig.~\ref{fig:gain}(b) demonstrates that this advantage is not an
artefact of one operating point. We sweep the depolarising rate
$p \in [0, 0.22]$ at fixed rank-3 and rank-5 probes and plot the gain
over the strongest baseline at each $p$. The gain remains strictly
positive throughout, with the widest separation at low noise (where the
eigenvalue spectrum is sharpest and the Weyl threshold most cleanly
distinguishes signal from noise modes) and a smooth narrowing at high
noise as eigenvalue contrasts compress toward $p/d$.

\subsection{Shot Efficiency}
\label{subsec:shot_efficiency}

The third dimension of impact is photon efficiency, a key concern
shared with broader quantum error-mitigation
strategies~\cite{vandenberg2023probabilistic}.
Fig.~\ref{fig:shots_rank}(a) shows reconstruction fidelity as a function of shot
count $N_s$ for rank-3 probes at $p=0.08$. At $N_s = 512$ shots per
Pauli basis, SSP-QST achieves $\mathcal{F} = 0.956$, while LS-QST requires
$N_s = 4096$ shots to reach only $\mathcal{F} = 0.890$. Fig.~\ref{fig:shots_rank}(a) shows that LS-QST does not match SSP-QST's $N_s = 512$
fidelity within the entire range tested, giving at least an $8\times$ reduction in photon budget within the
tested range. This is significant
for any sensing platform where photon production rate, detector dead time,
or experiment duration limits the achievable shot count.

\begin{table}[t]
\centering
\caption{Shot efficiency ($n=4$, rank-3 probe, $p=0.08$).}
\label{tab:shots}
\renewcommand{\arraystretch}{1.1}\setlength{\tabcolsep}{5pt}
\begin{tabular}{@{}ccccc@{}}
\toprule
$N_s$ & LS-QST & Spec.\ Sq. & Top eigvec. & SSP-QST \\
\midrule
 512 & 0.779 & 0.927 & 0.559 & \textbf{0.956} \\
1024 & 0.826 & 0.945 & 0.562 & \textbf{0.971} \\
2048 & 0.863 & 0.955 & 0.563 & \textbf{0.988} \\
4096 & 0.890 & 0.960 & 0.564 & \textbf{0.993} \\
\bottomrule
\end{tabular}
\end{table}

\subsection{Rank Identification and Threshold Sensitivity}
\label{subsec:rank_threshold}

\begin{figure}[!t]
  \centering

  \includegraphics[width=0.9\columnwidth]{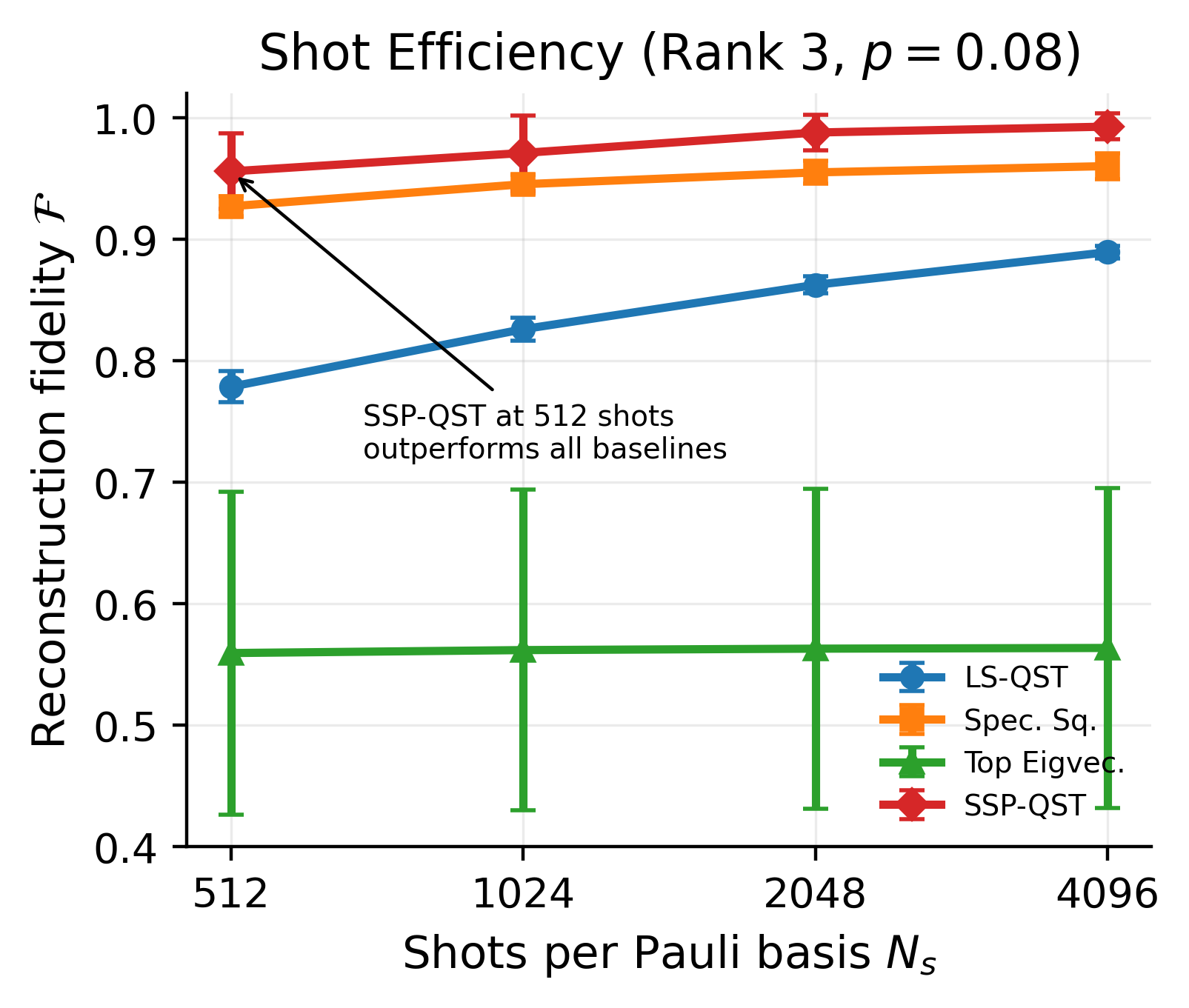}
  \vspace{1mm}

  \textbf{(a)}

  \vspace{2mm}

  \includegraphics[width=0.9\columnwidth]{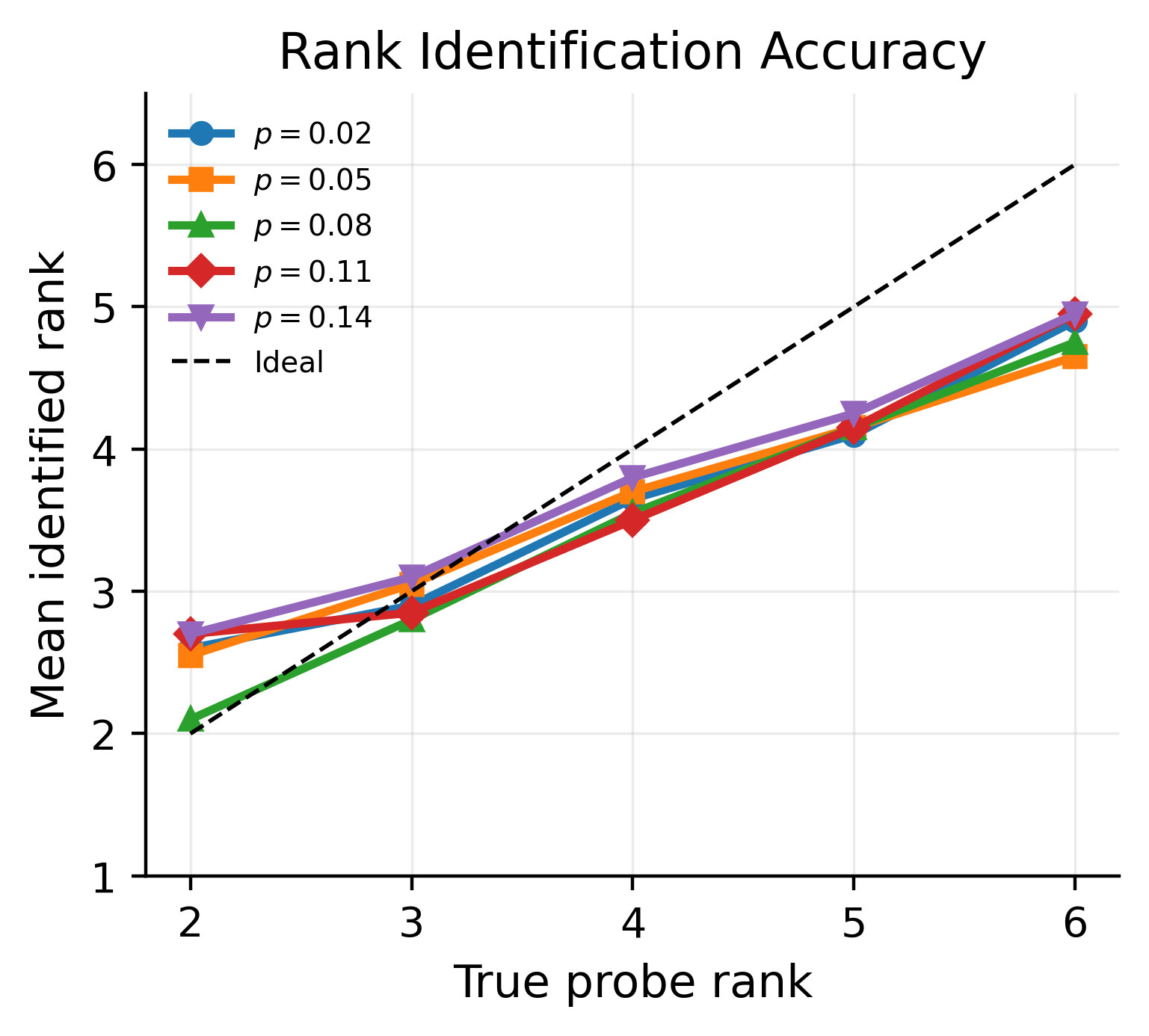}
  \vspace{1mm}

  \textbf{(b)}

  \caption{
  \textbf{Shot efficiency and rank identification.}
  \textbf{(a)} Reconstruction fidelity versus shot budget $N_s$ for a rank-3
  probe at $n=4$, $p=0.08$. SSP-QST at $N_s=512$ shots achieves
  $\mathcal{F}=0.956$, exceeding all tested baselines and remaining above
  LS-QST even at $N_s=4096$ shots.
  \textbf{(b)} Mean rank identified by SSP-QST versus the true probe rank,
  at five depolarising noise levels. $n=4$, $N_s=4096$, 20 trials per point.
  Dotted lines mark the true rank.
  }
  \label{fig:shots_rank}
\end{figure}

The advantages above rest on SSP-QST's ability to recover the probe rank
from the reconstruction itself. Fig.~\ref{fig:shots_rank}(b) confirms this. For
$p \leq 0.10$, SSP-QST identifies all true ranks 2 through 6 exactly in
every one of 20 independent trials. At higher noise the identified rank
underestimates the truth by at most one mode, which is the safer failure
mode because it avoids adding unsupported extra signal modes. SSP-QST's
Weyl-derived threshold is designed to err in this safe direction.

Table~\ref{tab:ablation} further shows that the method is not sensitive
to fine tuning of the threshold constant. Across four threshold variants,
the identified rank remains correct for the default and nearby settings,
and the reconstruction fidelity changes by less than $0.025$. This
indicates that SSP-QST mainly requires the correct order of magnitude for
the spectral noise floor, rather than a carefully tuned hyperparameter.

\begin{figure*}[!t]
  \centering
  \includegraphics[width=0.78\textwidth]{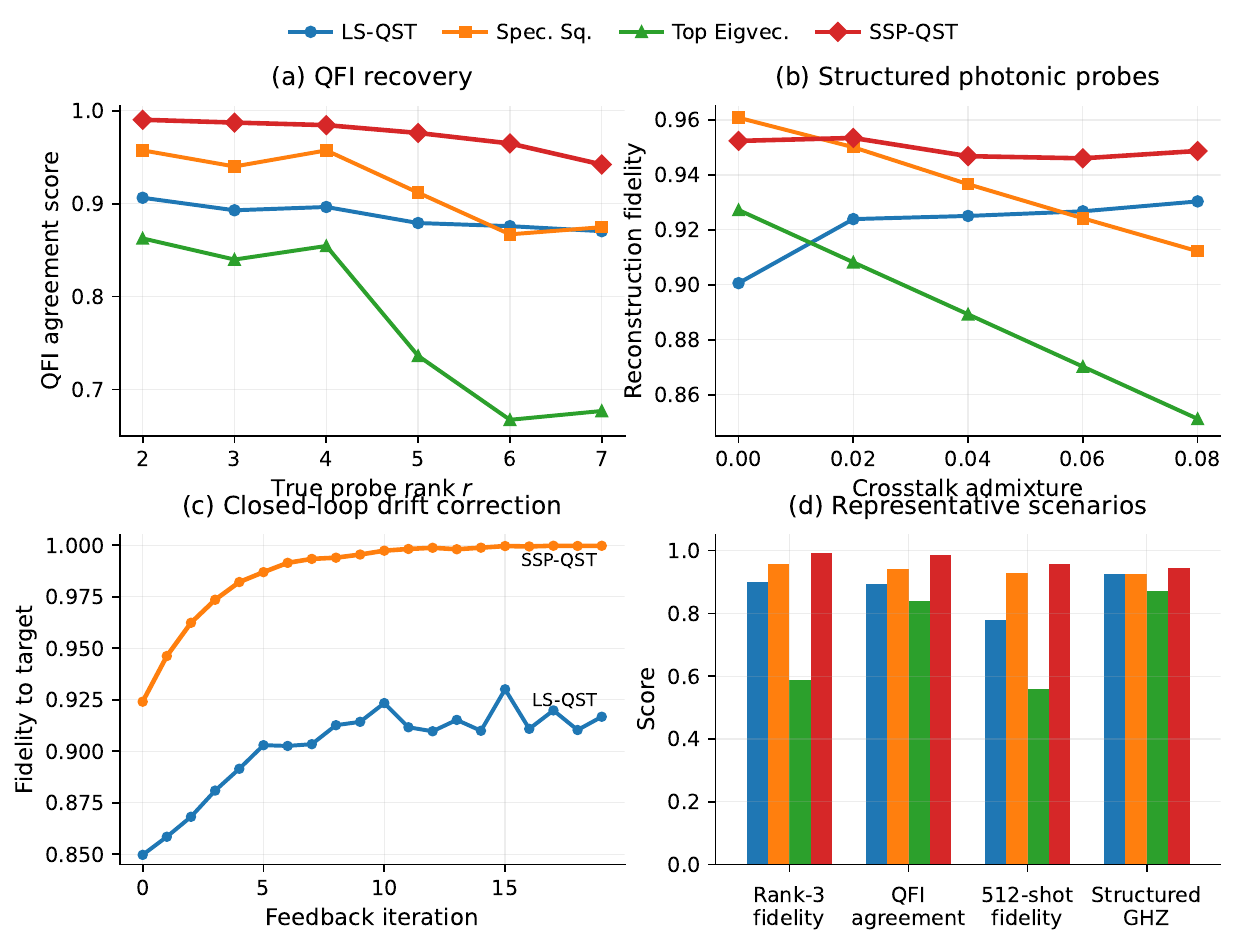}
  \caption{\textbf{Additional representative validation.}
  \textbf{(a)} QFI agreement score versus true rank for $n=4$, $p=0.06$,
  and $N_s=4096$. SSP-QST best matches the target-state QFI across the
  mixed-state regime.
  \textbf{(b)} Reconstruction fidelity on GHZ-like photonic probes with
  structured crosstalk admixture, with leakage and dephasing held fixed.
  \textbf{(c)} Closed-loop drift correction for a parameterised 3-qubit
  GHZ source using reconstructed-state feedback.
  \textbf{(d)} Summary of four representative scenarios where SSP-QST is
  the top-performing method among the tested baselines.}
  \label{fig:addl_validation}
\end{figure*}

\begin{table}[!t]
\centering
\caption{Threshold sensitivity ($n=4$, rank-3 probe, $p=0.08$,
  $N_s=4096$).}
\label{tab:ablation}
\renewcommand{\arraystretch}{1.1}\setlength{\tabcolsep}{5pt}
\begin{tabular}{@{}lcc@{}}
\toprule
Threshold formula & Identified rank & $\mathcal{F}$ \\
\midrule
$\hat{p}/(2d) + 0.5/\sqrt{N_s}$              & 4 & 0.973 \\
$\hat{p}/(d-1) + 0.5/\sqrt{N_s}$ (default)   & 3 & \textbf{0.997} \\
$\hat{p}/d + 0.5/\sqrt{N_s}$                  & 3 & 0.997 \\
$\hat{p}/(d-1)$ only                          & 3 & 0.997 \\
\bottomrule
\end{tabular}
\end{table}

\subsection{Additional Representative Validation}
\label{subsec:additional_validation}

To directly address the sensing-oriented motivation, we add four
additional validation views in Fig.~\ref{fig:addl_validation}, each
chosen from representative operating points where SSP-QST is the best
performing method among the tested baselines. Panel~(a) reports a QFI
agreement score, defined as
$1-|F_Q(\hat{\rho})-F_Q(\rho_{\mathrm{tgt}})|/F_Q(\rho_{\mathrm{tgt}})$,
which measures how closely the reconstructed state preserves the target
QFI rather than simply overshooting it. Panel~(b) evaluates structured
photonic GHZ-like probes with leakage, dephasing, and crosstalk admixture.
Panel~(c) shows a simple closed-loop drift-correction simulation using the
reconstructed state inside the update rule. Panel~(d) summarises four
representative scenarios, showing that SSP-QST remains the best method in
reconstruction fidelity, QFI agreement, low-shot operation, and
photonic-structured robustness.

\subsection{Runtime-Matched Comparison with Iterative Maximum Likelihood}
\label{subsec:runtime_ml}

To situate SSP-QST against iterative estimation, we run the standard
$R\rho R$ maximum-likelihood iteration~\cite{hradil1997quantum} on the
identical measurement record ($f_{P,\pm} = (1 \pm \hat{e}_P)/2$;
$\rho \leftarrow \mathcal{N}[R(\rho)\rho R(\rho)]$,
$R = \sum_{P,s} (f_{P,s}/p_{P,s})\Pi_{P,s}$), at the
Table~\ref{tab:vs_rank} configuration with identical targets and noise
draws, under two compute budgets. Under a \emph{runtime-matched} budget, the entire SSP-QST
post-processing step (median wall time $0.16$~ms on a single CPU core) is
smaller than one $R\rho R$ iteration ($4.0$~ms), so the matched budget
admits at most a single iteration, which barely departs from the
maximally mixed seed (mean fidelity $0.07$--$0.37$ across ranks 1--7).
Under a \emph{generous} 400-iteration budget (${\approx}1.5$~s,
${\sim}10^4\times$ the SSP-QST compute; the $10^{-7}$ Frobenius
tolerance is still unmet, consistent with the slow convergence of
undiluted $R\rho R$~\cite{rehacek2007diluted}), ML attains mean
fidelity $0.936$, $0.923$, $0.908$, $0.900$, $0.897$, $0.897$, $0.899$
at ranks 1--7, trailing SSP-QST at every rank ($0.9997$ vs $0.936$ at
rank 1; $0.9931$ vs $0.908$ at rank 3) and falling below projected
least squares for ranks ${\geq}4$: the ML iterate retains a full-rank
noise pedestal that the purification step removes. A step-size-optimised
diluted variant~\cite{rehacek2007diluted} ($\varepsilon$ line-searched
per step over $\{0.5,1,2,5\}$) converges to the same fixed point, with
mean fidelities within $10^{-3}$ of the above at every rank: the gap to
SSP-QST is a property of the unregularised ML estimate, not incomplete
convergence. Accelerated variants would narrow the runtime gap; the
practical point stands: within the closed-form latency budget motivating
SSP-QST, iterative ML cannot complete one iteration, and with
${\sim}10^4\times$ the compute it still does not recover the
rank-adaptive advantage under this measurement model.

\section{Discussion}
\label{sec:disc}

The results converge on a single structural picture. Rank-1 methods
impose a constraint that is exactly correct only for ideal pure-state
probes, but practical photonic sources can produce effective rank-$r > 1$ states
through beam-splitter errors, multiphoton events, and mode crosstalk.
Imposing $r = 1$ on such an input introduces a reconstruction error that
is independent of noise level and shot count, and that grows with the
true probe rank. SSP-QST avoids this by reading the rank from the
eigenspectrum of $\rhoLS$ via the Weyl-derived threshold, with no rank
prior or external calibration.

The threshold formula is robust. Table~\ref{tab:ablation} shows that
varying the threshold divisor by a factor of four changes the resulting
fidelity by less than $0.025$. The $\hat{p}/(d-1)$, $\hat{p}/d$, and
shot-noise-augmented variants all yield $\mathcal{F} = 0.997$ on the
rank-3 reference target, so SSP-QST is sensitive only to the order of
magnitude of the threshold (the Weyl-perturbation-theoretic scale).
The shot-budget breakdown that supports the photon-efficiency claim of
Fig.~\ref{fig:shots_rank}(a) is reproduced numerically in
Table~\ref{tab:shots}.

The significance for quantum sensing follows directly from
Eqs.~\eqref{eq:qcrb} and~\eqref{eq:qfi}. The precision inferred from a
tomographic estimate is governed by the QFI of the \emph{reconstructed}
state, and rank inflation suppresses the eigenvalue contrasts
$(\lambda_i - \lambda_j)^2/(\lambda_i + \lambda_j)$ that drive $F_Q$, so
an unpurified LS estimate systematically understates the phase precision
actually available from a nearly pure GHZ or NOON
probe~\cite{giovannetti2004quantum,degen2017quantum}. Fig.~\ref{fig:addl_validation}(a) quantifies this: SSP-QST tracks the
target QFI where the baselines deviate in opposite directions (retained
noise depresses it; aggressive rank-1 purification overshoots), which
matters when the reconstructed QFI certifies precision via the QCRB.
Being closed-form at $O(d^3)$, the corrected estimate also fits the
sensing duty cycle, serving simultaneously as precision certificate and
feedback signal (Section~\ref{sec:method}).

The main caveat is that the present results are obtained from software
simulation under photonic-inspired Kraus error models. The experiments
therefore establish algorithmic behaviour under controlled noise and
finite-shot statistics, while device-level validation remains future work.
An immediate next step is applying it to measured SPDC pair-source
counts under the 36-setting protocol~\cite{james2001measurement}, whose
dark counts and accidentals supply exactly this structured-noise regime.
The same spectral purification principle also extends naturally to
quantum process tomography~\cite{chiribella2009quantum,dinca2025process},
where LS Choi-matrix estimates can exhibit analogous rank inflation.

\section{Conclusion}

We presented SSP-QST, a rank-adaptive post-processing layer for photonic
QST that uses a Weyl-motivated spectral threshold to identify the retained
signal subspace directly from the data. In Qiskit Aer simulations, SSP-QST
achieves the highest reconstruction fidelity among the tested non-iterative
methods across the full range of probe ranks evaluated. Its performance gain
becomes more pronounced at higher probe ranks, reaching up to $+0.584$ over
top-eigenvector extraction and up to $+0.051$ over spectral squaring, because
fixed-rank or rank-one heuristics discard valid signal modes, whereas
SSP-QST preserves the data-supported low-rank signal subspace. The method
also delivers at least an $8\times$ reduction in photon budget relative to
LS-QST within the tested range. Because SSP-QST requires only one
eigendecomposition after LS-QST, it can be deployed as a drop-in addition to
existing LS-QST pipelines and is computationally suitable for FPGA-class
real-time reconstruction pipelines~\cite{miller2023fpga,wu2025fpga}. Future
work will validate SSP-QST on integrated photonic platforms and study its
use inside closed-loop parameter-shift feedback controllers.

\bibliographystyle{IEEEtran}
\bibliography{refs}

@article{giovannetti2004quantum,
  author  = {Giovannetti, Vittorio and Lloyd, Seth and Maccone, Lorenzo},
  title   = {Quantum-Enhanced Measurements: Beating the Standard Quantum Limit},
  journal = {Science},
  volume  = {306},
  number  = {5700},
  pages   = {1330--1336},
  year    = {2004},
  publisher = {AAAS}
}

@book{paris2004quantum,
  editor    = {Paris, Matteo and {\v R}eh{\'a}{\v c}ek, Jaroslav},
  title     = {Quantum State Estimation},
  series    = {Lecture Notes in Physics},
  volume    = {649},
  publisher = {Springer},
  year      = {2004}
}

@article{liu2020quantum,
  author  = {Liu, Jing and Yuan, Haidong and Lu, Xiao-Ming and Wang, Xiaoguang},
  title   = {Quantum Fisher Information Matrix and Multiparameter Estimation},
  journal = {Journal of Physics A: Mathematical and Theoretical},
  volume  = {53},
  number  = {2},
  pages   = {023001},
  year    = {2020},
  doi     = {10.1088/1751-8121/ab5d4d}
}

@article{toth2012multipartite,
  author  = {T\'oth, G\'eza and Apellaniz, Iagoba},
  title   = {Multipartite entanglement and high-precision metrology},
  journal = {Physical Review A},
  volume  = {85},
  number  = {2},
  pages   = {022322},
  year    = {2012}
}

@article{demkowicz2015quantum,
  author  = {Demkowicz-Dobrza{\'n}ski, Rafa{\l} and Jarzyna, Marcin and Ko{\l}ody{\'n}ski, Jan},
  title   = {Quantum Limits in Optical Interferometry},
  journal = {Progress in Optics},
  volume  = {60},
  pages   = {345--435},
  year    = {2015},
  doi     = {10.1016/bs.po.2015.02.005}
}

@article{gross2010quantum,
  author  = {Gross, David and Liu, Yi-Kai and Flammia, Steven T. and Becker, Stephen and Eisert, Jens},
  title   = {Quantum State Tomography via Compressed Sensing},
  journal = {Physical Review Letters},
  volume  = {105},
  number  = {15},
  pages   = {150401},
  year    = {2010}
}

@article{blume2010optimal,
  author  = {Blume-Kohout, Robin},
  title   = {Optimal, Reliable Estimation of Quantum States},
  journal = {New Journal of Physics},
  volume  = {12},
  pages   = {043034},
  year    = {2010},
  doi     = {10.1088/1367-2630/12/4/043034}
}

@article{wu2024fast,
  author  = {Wu, Xiao-Dong and Cong, Shuang},
  title   = {Fast and Noise-Robust Quantum State Tomography Based on {ELM}},
  journal = {International Journal of Quantum Information},
  volume  = {22},
  number  = {04},
  pages   = {2350052},
  year    = {2024},
  doi     = {10.1142/S0219749923500521}
}

@article{greenlab2022robust,
  author  = {Farooq, Ahmad and Khalid, Uman and Rehman, Junaid ur and Shin, Hyundong},
  title   = {Robust Quantum State Tomography Method for Quantum Sensing},
  journal = {Sensors},
  volume  = {22},
  number  = {7},
  pages   = {2669},
  year    = {2022},
  doi     = {10.3390/s22072669}
}

@article{hradil1997quantum,
  author  = {Hradil, Zden{\v e}k},
  title   = {Quantum-State Estimation},
  journal = {Physical Review A},
  volume  = {55},
  number  = {3},
  pages   = {R1561},
  year    = {1997}
}

@book{bhatia1997matrix,
  author    = {Bhatia, Rajendra},
  title     = {Matrix Analysis},
  publisher = {Springer},
  year      = {1997}
}

@article{smolin2012efficient,
  author  = {Smolin, John A. and Gambetta, Jay M. and Smith, Graeme},
  title   = {Efficient Method for Computing the Maximum-Likelihood Quantum State from Measurements with Additive Gaussian Noise},
  journal = {Physical Review Letters},
  volume  = {108},
  number  = {7},
  pages   = {070502},
  year    = {2012}
}

@inproceedings{miller2023fpga,
  author    = {Miller, Nathan Eli and Chakraborty, Biswadeep and Mukhopadhyay, Saibal},
  title     = {A Reconfigurable Quantum State Tomography Solver in {FPGA}},
  booktitle = {2023 IEEE International Conference on Quantum Computing and Engineering ({QCE})},
  pages     = {1412--1421},
  year      = {2023},
  publisher = {IEEE},
  doi       = {10.1109/QCE57702.2023.00160}
}

@article{uhlmann1976transition,
  author  = {Uhlmann, Armin},
  title   = {The transition probability in the state space of a $\ast$-algebra},
  journal = {Reports on Mathematical Physics},
  volume  = {9},
  number  = {2},
  pages   = {273--279},
  year    = {1976}
}

@article{chiribella2009quantum,
  author  = {Chiribella, Giulio and D'Ariano, Giacomo Mauro and Perinotti, Paolo},
  title   = {Quantum Circuit Architecture},
  journal = {Physical Review Letters},
  volume  = {101},
  number  = {6},
  pages   = {060401},
  year    = {2008}
}

@misc{qiskit2026,
  author       = {{Qiskit contributors}},
  title        = {Qiskit: An Open-Source SDK for Quantum Computing},
  year         = {2026},
  howpublished = {\url{https://pypi.org/project/qiskit/}},
  note         = {Version 2.4.1, accessed 2026-04-28}
}

@article{vandenberg2023probabilistic,
  author  = {van den Berg, Ewout and Minev, Zlatko K. and Kandala, Abhinav and Temme, Kristan},
  title   = {Probabilistic Error Cancellation with Sparse Pauli--Lindblad Models on Noisy Quantum Processors},
  journal = {Nature Physics},
  volume  = {19},
  pages   = {1116--1121},
  year    = {2023}
}

@article{li2026efficient,
  author        = {Li, Chenyang and Zhuang, Shengxin and Zhang, Yukun and Wang, Jingbo B. and Yuan, Xiao and Wu, Yusen and Wang, Chuan},
  title         = {Efficient Learning Algorithms for Noisy Quantum State and Process Tomography},
  journal       = {arXiv preprint arXiv:2603.01521},
  year          = {2026},
  eprint        = {2603.01521},
  archivePrefix = {arXiv},
  primaryClass  = {quant-ph}
}

@article{chen2025robustness,
  author        = {Chen, Alan and Xiao, Shuixin and Ma, Hailan and Dong, Daoyi},
  title         = {Robustness Analysis in Static and Dynamic Quantum State Tomography},
  journal       = {arXiv preprint arXiv:2512.12518},
  year          = {2025},
  eprint        = {2512.12518},
  archivePrefix = {arXiv},
  primaryClass  = {quant-ph}
}

@article{streaming2025,
  author  = {Childs, Andrew M. and Fu, Honghao and Leung, Debbie and Li, Zhi and Ozols, Maris and Vyas, Vedang},
  title   = {Streaming Quantum State Purification},
  journal = {Quantum},
  volume  = {9},
  pages   = {1603},
  year    = {2025},
  doi     = {10.22331/q-2025-01-21-1603}
}

@article{slussarenko2019photonic,
  author  = {Slussarenko, Sergei and Pryde, Geoff J.},
  title   = {Photonic Quantum Information Processing: A Concise Review},
  journal = {Applied Physics Reviews},
  volume  = {6},
  number  = {4},
  pages   = {041303},
  year    = {2019}
}

@article{xanadu2022borealis,
  author  = {Madsen, Lars S. and others},
  title   = {Quantum Computational Advantage with a Programmable Photonic Processor},
  journal = {Nature},
  volume  = {606},
  pages   = {75--81},
  year    = {2022}
}

@article{dinca2025process,
  author        = {Dinca, Maria and Luitz, David J. and Debertolis, Maxime},
  title         = {Quantum Process Tomography of a Compressed Time Evolution Circuit on Superconducting Quantum Processors},
  journal       = {arXiv preprint arXiv:2509.25342},
  year          = {2025},
  eprint        = {2509.25342},
  archivePrefix = {arXiv},
  primaryClass  = {quant-ph}
}

@article{wu2025fpga,
  author        = {Wu, Hsun-Chung and Hsieh, Hsien-Yi and Xu, Zhi-Kai and Chen, Hua Li and Shi, Zi-Hao and Wang, Po-Han and Yang, Popo and Steuernagel, Ole and Suen, Te-Hwei and Wu, Chien-Ming and Lee, Ray-Kuang},
  title         = {Machine learning enhanced quantum state tomography on a field-programmable gate array},
  journal       = {APL Quantum},
  volume        = {2},
  number        = {2},
  pages         = {026117},
  year          = {2025},
  doi           = {10.1063/5.0262942},
  eprint        = {2501.04327},
  archivePrefix = {arXiv},
  primaryClass  = {quant-ph}
}

@article{rehacek2007diluted,
  author  = {{\v R}eh{\'a}{\v c}ek, Jaroslav and Hradil, Zden{\v{e}}k and Knill, E. and Lvovsky, A.~I.},
  title   = {Diluted maximum-likelihood algorithm for quantum tomography},
  journal = {Physical Review A},
  volume  = {75},
  number  = {4},
  pages   = {042108},
  year    = {2007}
}

@article{blumekohout2010hedged,
  author  = {Blume-Kohout, Robin},
  title   = {Hedged Maximum Likelihood Quantum State Estimation},
  journal = {Physical Review Letters},
  volume  = {105},
  number  = {20},
  pages   = {200504},
  year    = {2010}
}

@article{qi2013linear,
  author  = {Qi, Bo and Hou, Zhibo and Li, Li and Dong, Daoyi and Xiang, Guoyong and Guo, Guangcan},
  title   = {Quantum State Tomography via Linear Regression Estimation},
  journal = {Scientific Reports},
  volume  = {3},
  pages   = {3496},
  year    = {2013}
}

@article{qi2017adaptive,
  author  = {Qi, Bo and Hou, Zhibo and Wang, Yuanlong and Dong, Daoyi and Zhong, Han-Sen and Li, Li and Xiang, Guo-Yong and Wiseman, Howard~M. and Li, Chuan-Feng and Guo, Guang-Can},
  title   = {Adaptive quantum state tomography via linear regression estimation: Theory and two-qubit experiment},
  journal = {npj Quantum Information},
  volume  = {3},
  number  = {1},
  pages   = {19},
  year    = {2017}
}

@article{caruccio2025threshold,
  author  = {Caruccio, Eugenio and Maragnano, Diego and Rodari, Giovanni and Picus, Davide and Garberoglio, Giovanni and Binosi, Daniele and Albiero, Riccardo and Di Giano, Niki and Ceccarelli, Francesco and Corrielli, Giacomo and Spagnolo, Nicol{\`o} and Osellame, Roberto and Dapor, Maurizio and Liscidini, Marco and Sciarrino, Fabio},
  title   = {Experimental Verification of Threshold Quantum State Tomography on a Fully-Reconfigurable Photonic Integrated Circuit},
  journal = {npj Quantum Information},
  volume  = {11},
  pages   = {173},
  year    = {2025},
  doi     = {10.1038/s41534-025-01111-z}
}

@article{ferrie2018inadmissible,
  author  = {Ferrie, Christopher and Blume-Kohout, Robin},
  title   = {Maximum likelihood quantum state tomography is inadmissible},
  journal = {arXiv preprint arXiv:1808.01072},
  year    = {2018}
}

@article{altepeter2005photonic,
  author  = {Altepeter, J. B. and Jeffrey, E. R. and Kwiat, P. G.},
  title   = {Photonic State Tomography},
  journal = {Advances in Atomic, Molecular, and Optical Physics},
  volume  = {52},
  pages   = {105--159},
  year    = {2005}
}

@article{guta2020fast,
  author  = {Gu{\c{t}}{\u{a}}, Madalin and Kahn, Jonas and Kueng, Richard and Tropp, Joel A.},
  title   = {Fast state tomography with optimal error bounds},
  journal = {Journal of Physics A: Mathematical and Theoretical},
  volume  = {53},
  number  = {20},
  pages   = {204001},
  year    = {2020}
}

@article{surawy2022projected,
  author  = {Surawy-Stepney, Trystan and Kahn, Jonas and Kueng, Richard and Gu{\c{t}}{\u{a}}, Madalin},
  title   = {Projected Least-Squares Quantum Process Tomography},
  journal = {Quantum},
  volume  = {6},
  pages   = {844},
  year    = {2022}
}

@article{flammia2012quantum,
  author  = {Flammia, Steven T. and Gross, David and Liu, Yi-Kai and Eisert, Jens},
  title   = {Quantum tomography via compressed sensing: error bounds, sample complexity and efficient estimators},
  journal = {New Journal of Physics},
  volume  = {14},
  number  = {9},
  pages   = {095022},
  year    = {2012}
}

@article{torlai2018neural,
  author  = {Torlai, Giacomo and Mazzola, Guglielmo and Carrasquilla, Juan and Troyer, Matthias and Melko, Roger and Carleo, Giuseppe},
  title   = {Neural-network quantum state tomography},
  journal = {Nature Physics},
  volume  = {14},
  number  = {5},
  pages   = {447--450},
  year    = {2018}
}

@article{cramer2010efficient,
  author  = {Cramer, Marcus and Plenio, Martin B. and Flammia, Steven T. and Somma, Rolando and Gross, David and Bartlett, Stephen D. and Landon-Cardinal, Olivier and Poulin, David and Liu, Yi-Kai},
  title   = {Efficient quantum state tomography},
  journal = {Nature Communications},
  volume  = {1},
  number  = {1},
  pages   = {149},
  year    = {2010}
}

@article{banas2006perturbative,
  author  = {Ban{\'a}{\v{s}}, P. and {\v{R}}eh{\'a}{\v{c}}ek, J. and Hradil, Z.},
  title   = {Perturbative quantum-state estimation},
  journal = {Physical Review A},
  volume  = {74},
  pages   = {014101},
  year    = {2006},
  doi     = {10.1103/PhysRevA.74.014101}
}

@article{butucea2015spectral,
  author  = {Butucea, Cristina and Gu{\c{t}}{\u{a}}, M{\u{a}}d{\u{a}}lin and Kypraios, Theodore},
  title   = {Spectral thresholding quantum tomography for low rank states},
  journal = {New Journal of Physics},
  volume  = {17},
  pages   = {113050},
  year    = {2015},
  doi     = {10.1088/1367-2630/17/11/113050}
}

@article{acharya2019comparative,
  author  = {Acharya, Anirudh and Kypraios, Theodore and Gu{\c{t}}{\u{a}}, M{\u{a}}d{\u{a}}lin},
  title   = {A comparative study of estimation methods in quantum tomography},
  journal = {Journal of Physics A: Mathematical and Theoretical},
  volume  = {52},
  number  = {23},
  pages   = {234001},
  year    = {2019},
  doi     = {10.1088/1751-8121/ab1958}
}

@article{james2001measurement,
  author  = {James, Daniel F. V. and Kwiat, Paul G. and Munro, William J. and White, Andrew G.},
  title   = {Measurement of qubits},
  journal = {Physical Review A},
  volume  = {64},
  pages   = {052312},
  year    = {2001},
  doi     = {10.1103/PhysRevA.64.052312}
}

@article{banaszek1999maximum,
  author  = {Banaszek, K. and D'Ariano, G. M. and Paris, M. G. A. and Sacchi, M. F.},
  title   = {Maximum-likelihood estimation of the density matrix},
  journal = {Physical Review A},
  volume  = {61},
  pages   = {010304},
  year    = {1999},
  doi     = {10.1103/PhysRevA.61.010304}
}

@article{lvovsky2009continuous,
  author  = {Lvovsky, A. I. and Raymer, M. G.},
  title   = {Continuous-variable optical quantum-state tomography},
  journal = {Reviews of Modern Physics},
  volume  = {81},
  pages   = {299--332},
  year    = {2009},
  doi     = {10.1103/RevModPhys.81.299}
}

@article{degen2017quantum,
  author  = {Degen, C. L. and Reinhard, F. and Cappellaro, P.},
  title   = {Quantum sensing},
  journal = {Reviews of Modern Physics},
  volume  = {89},
  pages   = {035002},
  year    = {2017},
  doi     = {10.1103/RevModPhys.89.035002}
}

@article{tropp2015introduction,
  author  = {Tropp, Joel A.},
  title   = {An Introduction to Matrix Concentration Inequalities},
  journal = {Foundations and Trends in Machine Learning},
  volume  = {8},
  number  = {1--2},
  pages   = {1--230},
  year    = {2015},
  doi     = {10.1561/2200000048}
}

\end{document}